\documentclass[12pt,preprint]{aastex}
\usepackage{epsf}
\usepackage{graphicx}

\newcommand{\etal}{{et al.~}}
\newcommand {\apgt} {\ {\raise-.5ex\hbox{$\buildrel>\over\sim$}}\ }
\newcommand {\aplt} {\ {\raise-.5ex\hbox{$\buildrel<\over\sim$}}\ }

\begin{document}

\pagenumbering{arabic}
\title{Spitzer Space Telescope Observations of the Aftermath of 
Microlensing Event MACHO-LMC-5}
\author{Hien T.\ Nguyen}
\affil{Jet Propulsion Laboratory, California Institute of Technology,
4800 Oak Grove Drive, Pasadena, CA 91109}
\email{Hien.T.Nguyen@jpl.nasa.gov}
\author{Nitya Kallivayalil}
\affil{Department of Physics and Astronomy, University
of Pennsylvania, David Rittenhouse Laboratory,
209 S.\ 33rd St., Philadelphia, PA 19104}
\email{nitya@student.physics.upenn.edu}
\author{Michael W.\ Werner}
\affil{Jet Propulsion Laboratory, California Institute of Technology
4800 Oak Grove Drive, Pasadena, CA 91109}
\email{mwerner@sirtfweb.jpl.nasa.gov}
\author{Charles Alcock}
\affil{Department of Physics and Astronomy, University
of Pennsylvania, David Rittenhouse Laboratory,
209 S.\ 33rd St., Philadelphia, PA 19104}
\email{alcock@hep.upenn.edu}
\author{Brian M.\ Patten}
\affil{Harvard-Smithsonian Center for Astrophysics, 60 Garden St. 
Cambridge, MA 02138-1516} 
\email{bpatten@cfa.harvard.edu}
\and
\author{Daniel Stern}
\affil{Jet Propulsion Laboratory, California Institute of Technology,
4800 Oak Grove Drive, Pasadena, CA 91109}
\email{stern@zwolfkinder.jpl.nasa.gov}

\begin{abstract}
We have carried out photometry of the microlensing event MACHO-LMC-5 
with Spitzer's IRAC ten years after the magnification of the LMC source 
star was recorded. This event is unique in the annals of 
gravitational microlensing: the lensing star 
itself has been observed using HST (once with WFPC2 and twice with ACS/HRC). 
Since the separation between the source and lens at the epoch of the Spitzer 
observations was $\sim0.24''$, the two stars cannot be resolved in the 
Spitzer images. However, the IRAC photometry clearly establishes that the 
lens is a M5 dwarf star from its infrared excess, which in turn yields
a mass of $\sim0.2 M_{\odot}$.  This demonstrates the 
potential of Spitzer to detect the lenses in other gravitational 
microlensing events.

\end{abstract}

\section{Introduction}

It is over a decade since gravitational microlensing was first clearly 
detected (Alcock \etal 1993), and there has been enormous progress since then 
(Alcock \etal 2000a; Afonso \etal 2003). Over 1,000 events have been recorded, most towards the 
Galactic bulge; the total towards the Large and Small 
Magellanic Clouds is $\sim25$ (Alcock \etal 2000a; Afonso \etal 2003). The Magellanic Cloud 
events are of great importance because they probe the contribution of 
MACHOs (Massive Compact Halo Objects) to the dark matter in the halo 
of the Milky Way (Paczynski 1986). 
The measured microlensing event rate towards the Large Magellanic 
Cloud exceeds that expected from previously known stellar populations 
(Alcock \etal 2000a). The interpretation of this excess of events remains 
controversial, with four potential explanations dominating the discussion:

\begin{enumerate}

\item The lenses belong to the halo of the Milky Way.
In this case the MACHO Project concluded that objects of
mass $\sim 0.5 M_{\odot}$ comprise about 20\% of the dark halo (Alcock
\etal 2000a).

\item The lenses belong to some previously undetected population that
belongs to the LMC (Sahu 1994; see also Zhao 1998).
Models of the LMC do not generally produce sufficient microlensing
event rates to account for the data (Gould 1995, 1998; Alves \&
Nelson 2000; Gyuk, Dalal, \& Griest 2000; van der Marel \etal 2002). 
Tidal debris that is not in dynamical equilibrium with the LMC
has also been discussed (Weinberg \& Nikolaev 2001; Zhao \etal 2003). 
Efforts to detect various models for
these putative populations have been unsuccessful (Alcock \etal 2001a;
Alves 2004).

\item The lenses belong to some previously undetected dwarf galaxy that lies
between us and the LMC (Zaritsky \& Lin 1997; Zaritsky \etal 1999).
This is now considered very improbable (Gould 1999).

\item The lenses belong to a previously undetected component of the
disk of the Milky Way (Alcock \etal 2000a; note that the {\it known} components
cannot account for the microlensing event rate), or to an undetected 
disk-like structure (Gates \& Gyuk 2001).
This suggestion is less popular
than (1) and (2), but does receive some support in the story of
event MACHO-LMC-5.

\end{enumerate}

The truth, of course, may be a combination of these,
and there is great interest in elucidating the true nature of
these events (Gates, Gyuk, \& Turner 1996; 
Kerins \& Evans 1999; Green \& Jedamzik 2002;
Jetzer, Mancini, \& Scarpetta 2002). 
It is in this context that the special importance of the
MACHO-LMC-5 is manifest. 
During a 76 day period in 1993, the brightness of MACHO-LMC-5 increased by a
factor of 47 due to the passage of a foreground object (the lens) 
close to our line of sight to the source. The lens and source are now
separated by $\sim0.24''$. MACHO-LMC-5
is the only event in history for which both the source star
{\it{and the lens}} have been detected (Alcock \etal 2001b). In the original
report on the HST WFPC2 image showing the lens,
the observed source-lens relative proper motion was strikingly
consistent with the motion determined by the fit to the microlensing
event itself. The mass of and distance to the lens were estimated
from the fit to the microlensing data, and separately from spectra
and the photometric data; these two independent estimates appeared
to differ significantly. Gould (2004) suggested that a phenomenon
called ``jerk parallax'' was responsible for this discrepancy. 
Microlensing parallax
is the asymmetry in a magnification event induced by the acceleration of the 
Earth around the Sun. ``Jerk'' is the time derivative of this acceleration.
Gould's interpretation was
confirmed by Drake, Cook \& Keller 2004 using new images of the system taken
with HST's ACS/HRC. They concluded that the lens is very probably
a dwarf M5 star at a distance of $\approx 600$pc from the sun.

We have begun a program of deep photometry of lensed LMC stars using
the Spitzer Space Telescope. We start with MACHO-LMC-5 because of its
great importance, because it should be readily detected in Spitzer's
IRAC bands, and because we plan to use these observations as a model
for further analysis of other LMC events. 
In most cases the lens
cannot be resolved from the source, even with HST resolution
(Alcock \etal 2001a), but a cool stellar lens in a Galactic disk population
may show up as an infrared excess (von Hippel \etal 2003). 
MACHO-LMC-5
is indeed detected, as we report here. Observations of additional LMC
microlensing source stars will be presented in a later paper.

We present data on the MACHO-LMC-5 system in which we have detected
the lens with Spitzer at wavelengths out to the 8$\mu$m band. We establish
photometrically that the lens is a late M star and also demonstrate
the utility of Spitzer observations for characterizing the lensing
population. In this paper, we will use the phrase  ``MACHO-LMC-5''  to
refer to the combination of the target star and the lens, which cannot
\
be spatially separated by Spitzer\footnote{The full-width at
  half-maximum of the PSF is  $\sim1.8''$. }.  ``LMC-5 lens''  refers to the
foreground cold object which produced the microlensing, and  ``LMC-5
source''  to the background LMC star which was magnified.

\section{Observations}

MACHO-LMC-5 was observed with the IRAC near/mid-infrared camera on
board Spitzer at UT 2003 December 05.
IRAC is a four-channel camera consisting of two pairs of
$256\times256$ pixel InSb and Si:As IBC detectors to provide
simultaneous images at 3.6, 4.5, 5.8, and
8$\mu$m. Two adjacent $5.12\times5.12$ arcmin 
fields of view are viewed by the four channels in pairs (3.6 and
5.8$\mu$m; 4.5 and 8$\mu$m) \footnote{IRAC 
3.6 and 4.5$\mu$m bands have effective wavelengths
similar to the widely used L and M filters respectively (Fazio et al. 
2004 this issue).}. The fields of view were centered on 05h16m40s,
-70d29m04s (J2000), which is the position of the LMC-5 event. 
The target area was imaged using a 12-position
    Reuleaux triangle dither pattern,  at approximately 7 to 8 arcseconds
    per step,  with two 30-second frametime exposures
    made at each dither position. The net result was 720 s of
integration on MACHO-LMC-5 in each IRAC band.  Additional 
observations of MACHO-LMC-5 with the MIPS 24$\mu$m band are planned for
April 2004. 

\section{Data Reduction and Analysis}

The individual frames were processed (to remove cosmic rays and
artifacts, primarily a streak due to a bright star just to the east of
MACHO-LMC-5; the streak was particularly noticeable at 5.8$\micron$ and
8$\micron$, where it interfered with our ability to obtain accurate
photometry of MACHO-LMC-5), and co-added to produce the images used 
for astrometry and photometry. The co-added frames have pixel size 0.6'',
half the native pixel scale for IRAC. 

Figure 1 shows the entire processed image of this field taken in the
4.5$\mu$m band.
Figure 2 shows six images, each of the same small
region including MACHO-LMC-5: (a) the MACHO Project $R$-band image, (b)
$I$-band with the HST WFPC2 imager, (c) 3.6$\mu$m IRAC image, (d)
4.5$\mu$m IRAC image, (e) 5.8$\mu$m IRAC image, and (f) 8$\mu$m IRAC
image. The lens and source are clearly resolved in the HST image, but
unresolved in the MACHO and IRAC images. Note that the complex of
stars around MACHO-LMC-5 seen in the MACHO image is clearly seen in
the Spitzer images, particularly at 3.6 and 4.5$\mu$m.

The MACHO-LMC-5 complex is the Eastern member of a pair of objects
separated by $\sim$2'' roughly in the E-W direction. The Western
member of the pair is the star shown to the right of MACHO-LMC-5 in
the HST image. This star is considerably brighter than the MACHO-LMC-5
complex in both $R$-band and $I$-band, but in the IRAC images MACHO-LMC-5
is about as bright as this star. This can be attributed to the fact
that the lens, which is substantially cooler than the source star,
dominates the flux at longer wavelengths.

\textit{Astrometry.} In order to identify MACHO-LMC-5 in our IRAC
data, we identified the region of interest using the lower resolution
MACHO-$R$-band image and then made use of HST $I$-band data on MACHO-LMC-5
(Figure 2(b)). The HST data does not suffer from confusion, and all
the relevant stars are well isolated. In addition, we know which the
lens, the source, and the field star are in the
HST image. We ran DAOPHOT on both the IRAC and HST fields
to get astrometric centroids for the stars. For each data set (IRAC
and HST) we then computed both the distance between MACHO-LMC-5 and a
bright reference star present in both the HST and the IRAC images,
and the distance between this bright star and the field
star. The ratio of distances determined for the two
separate images agreed to within a few percent, allowing us to
identify MACHO-LMC-5 and its closest neighbor with confidence in the
IRAC images.

\textit{Photometry.} Photometry was done on the cleaned, 
co-added images using DAOPHOT. In these mosaics the individual
frames in the dithers were 
coadded in sky coordinates with the SSC mosaicer software using 
0.6'' pixels, half the native pixel scale for IRAC. The PSF radius 
was set to 6''  (DAOPHOT
keyword \texttt{psfrad} $=$ 10 pixels), and the fit radius was set to
3.6''  (DAOPHOT keyword \texttt{fitrad} $=$ 6 pixels). Aperture
corrections were performed for these source apertures (3 pixels in the
native IRAC pixel scale). Magnitude zero
points were determined using data provided by the Spitzer Science
Center. The Jy to 0 mag conversions used were 277.5, 179.5, 116.6 and
63.1 for the 3.6,
4.5, 5.8 \& 8$\mu$m bands respectively. Table 1 shows all the 
photometric data available for MACHO-LMC-5 
as well as photometry for two late-type dwarfs used for comparison 
(Bessell 1991, Leggett 1992, Patten et al. 2004).

\section{Discussion and Conclusions}
The images and photometry in Table 1 clearly show that MACHO-LMC-5
exhibits substantial infrared excess. This was already apparent in the
separate $V-I$ colors for the source and the lens from the HST data. 
Given the faintness of the LMC-5 source star as seen in the 
  HST images and the likelihood that it is an early G-type
  star, based on its $V-I$ color (Alcock et al. 2001b), 
we estimate that the source star will contribute 
  less than 10\% of the flux of the combined MACHO-LMC-5 within the   
  IRAC bandpasses.  Therefore,
  within the errors on the photometry, MACHO-LMC-5 has mid-infrared
  colors consistent with a late-M dwarf or possibly an early-L dwarf
  (Patten \etal 2004).

The jerk parallax fit (Drake, Cook, \& Keller 2004) 
yields a distance of $578\pm65$ pc, 
equivalent to a distance modulus of $8.81\pm0.2$ for the lens. The 
corresponding absolute
magnitudes of the lens in IRAC bands are given in Table 2. 
 Table 2 also lists absolute
magnitudes from IRAC photometry of GJ1156 (M5.0V) and LHS3003 (M7.0V),
both of which have Hipparcos trigonometric parallaxes (Patten \etal 2004). 
Clearly the M5.0 V star is the best match for MACHO-LMC-5. Because the absolute magnitudes for low-mass stars fall off rapidly 
for later spectral types, this eliminates the consideration of L 
dwarfs as possibilities for the color match.  Likewise, the lens 
can be no earlier spectral type than M4.5 V.  Therefore, the IRAC 
photometry shows that the lens is an M5 dwarf and thus has a mass 
of $\sim0.2 M_{\odot}$ (Henry \& McCarthy 1993, Delfosse \etal 2000). 
The characterization of this spectacular event is 
now complete.  


The lens in MACHO-LMC-5 
belongs to the disk of the Milky Way.  The distance and space motions are
consistent with the thick disk, as noted by Drake, Cook \& Keller (2004).
Since a fraction of all microlensing candidates are
expected to come
from  the disk, we cannot say anything conclusive about the
nature of the microlenses in general with just this one
event. Thus, while MACHO-LMC-5 \textit{might} be an example of the
fourth possibility listed in the Introduction, \textit{on its own}
it does not suggest the existence of a previously unknown component.
However,
the clear detection of the infrared excess in MACHO-LMC-5 demonstrates
the capability of Spitzer to detect cool, low mass stellar
lenses as candidates for the other microlensing events if this is
indeed the case. Previous searches for infrared excesses have
yielded only upper limits (von Hippel \etal
2003). The sensitivity and
mid-infrared spectral coverage of Spitzer will allow much more
sensitive searches for the lenses corresponding to other microlensing
events in the future.

\begin{deluxetable}{lcccccccc}
\tabletypesize{\scriptsize}
\tablewidth{0pt}
\tablecolumns{8}
\tablecaption{The photometry}
\tablehead{
\colhead{Star} & \colhead{$V$} & \colhead{$(V - I)_c$\tablenotemark{\dag}} & \colhead{$3.6\micron$}
& \colhead{$4.5\micron$} &
\colhead{$5.8\micron$} & \colhead{$8\micron$} & 
\colhead{$V-3.6\micron$}}
\startdata
LMC-5 source & $21.02 \pm 0.06$ & $0.68 \pm 0.09$ & -  &  - & -
& - & - \\
LMC-5 lens & $22.67\pm0.10$ & $3.18\pm0.11$ & - & 
 - & - & - & - \\
MACHO-LMC-5 & $20.80\pm0.12$ & $1.72\pm0.14$ & $16.33\pm0.03$\tablenotemark{\P} & 
$16.47\pm0.04$ &
$16.70\pm0.15$ & $16.83\pm0.29$  & $4.47\pm0.12$ \\
GJ1156 & $13.80$ & $3.45$ & $7.18\pm0.02$ & $7.17\pm0.01$ & $7.13\pm0.03$ & $7.07\pm0.01$  & $6.61\pm0.02$ \\
LHS3003 & $17.05$ & $4.52$ & $8.42\pm0.03$ & $8.50\pm0.01$ & $8.42\pm0.01$ &
$8.33\pm0.01$ & $8.63\pm0.03$ \\
\enddata
\tablenotetext{\dag}{the c refers to the Cousins system.}
\tablenotetext{\P}{all error estimates quoted are for 
relative photometric errors. The absolute calibration of IRAC 
is believed to be accurate to 10\% at this time.}
\end{deluxetable}

\begin{deluxetable}{lccccc}
\tabletypesize{\small}
\tablewidth{0pt}
\tablecolumns{5}
\tablecaption{Absolute Magnitudes}
\tablehead{
\colhead{Star}  & \colhead{$M_{3.6\micron}$} & 
\colhead{$M_{4.5\micron}$} & \colhead{$M_{5.8\micron}$} &
\colhead{$M_{8\micron}$}}
\startdata
GJ1156 (M5.0V) & $8.11\pm0.10$ & $8.09\pm0.10$ & $8.05\pm0.10$ & $7.99\pm0.10$\\
LHS3003 (M7.0V) & $9.48\pm0.10$ & $9.56\pm0.10$ & $9.48\pm0.10$ & $9.39\pm0.10$\\
MACHO-LMC-5 &  $7.52\pm0.23$ & $7.66\pm0.23$ & $7.89\pm0.27$ 
& $8.02\pm0.37$\\
\enddata
\end{deluxetable}

\begin{figure}
\plotone{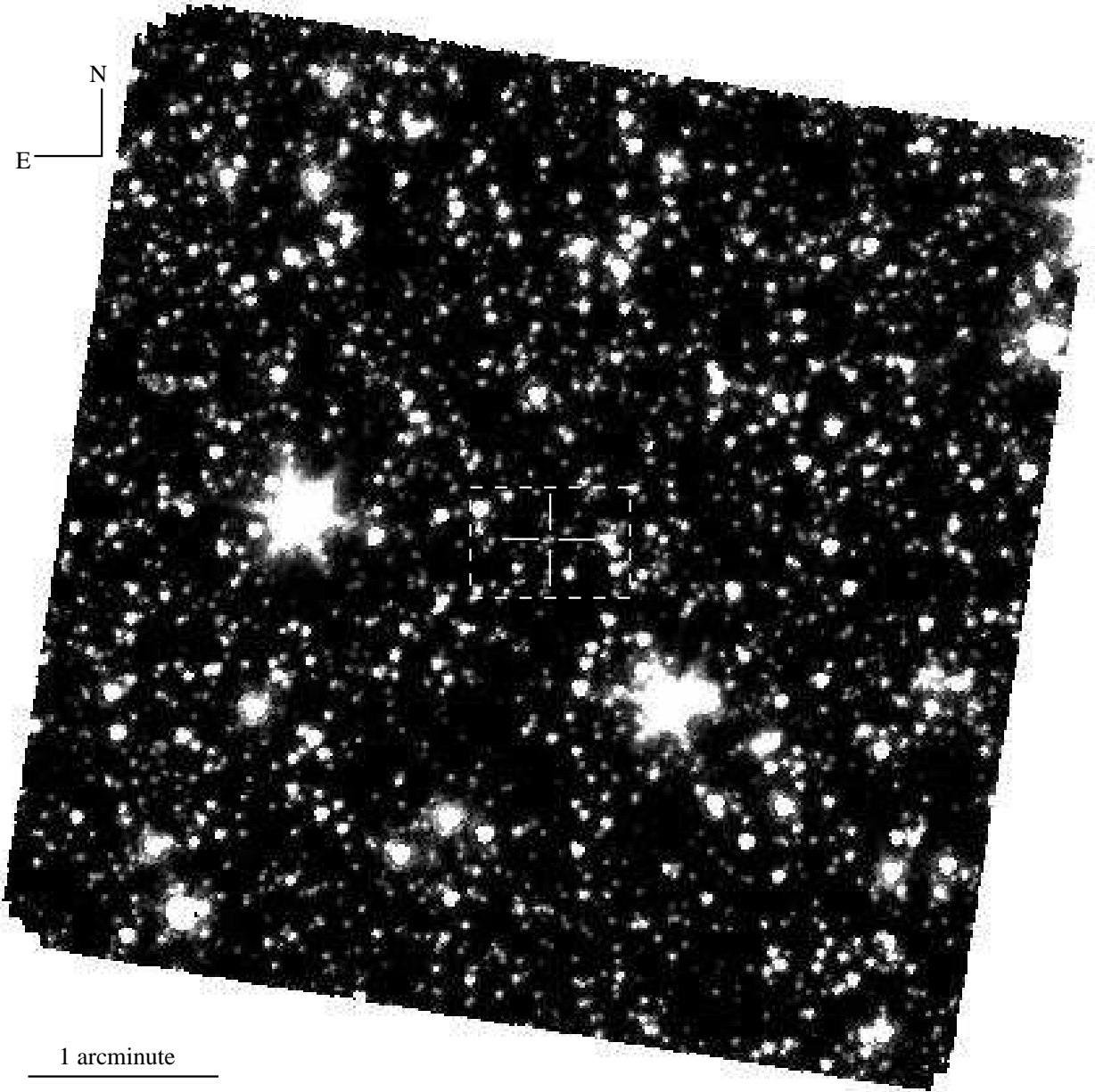}
\caption{A full image of the MACHO LMC-5 field taken with the IRAC
  4.5$\micron$ band. The boxed region highlights the MACHO-LMC-5
    (left) \& field star (right) pair.}
\end{figure}

\begin{figure}
\plotone{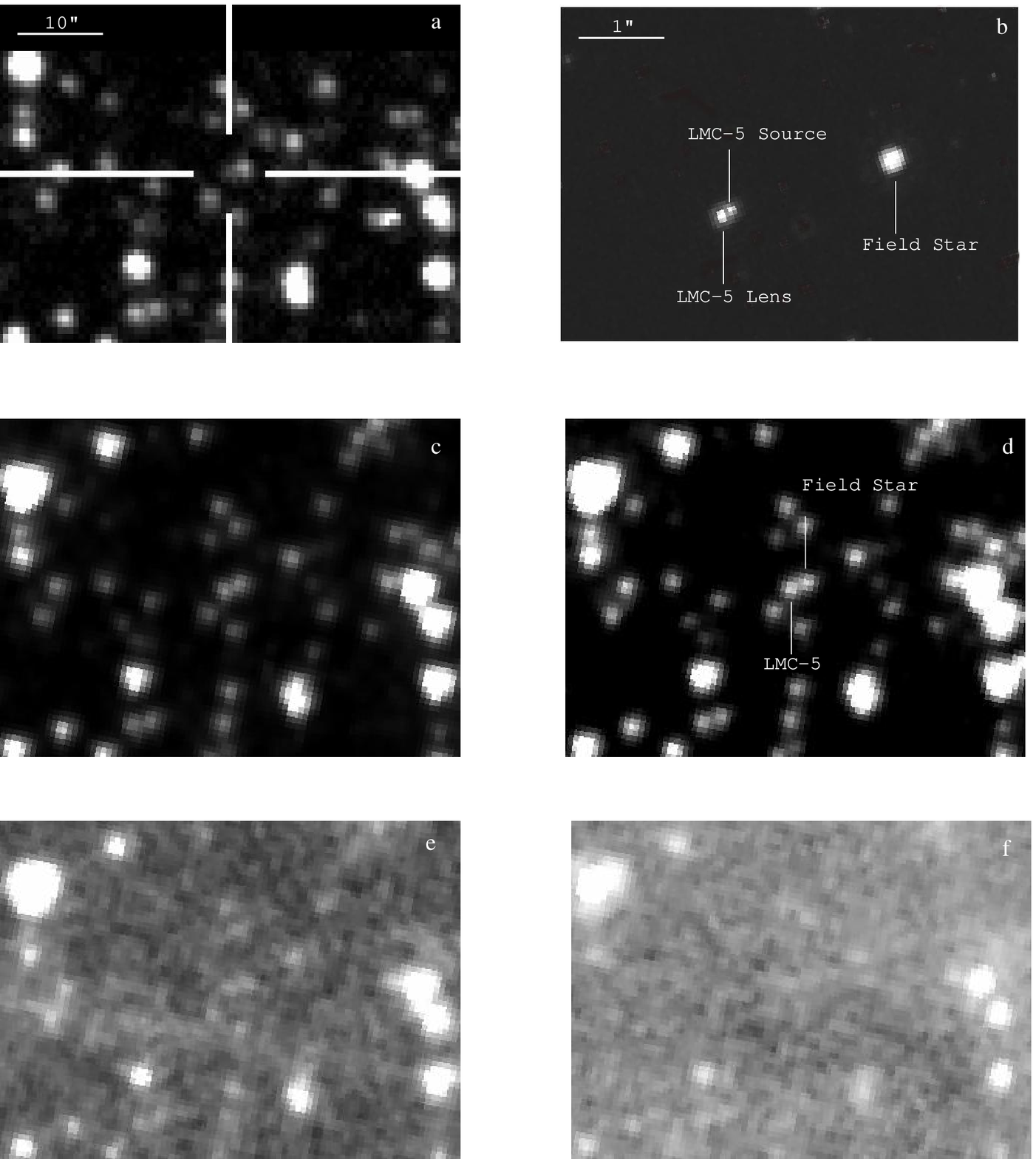}
\caption{\scriptsize{(a) Aftermath of 
MACHO-LMC-5 image in $R$-band (Alcock \etal 2001).
(b) High resolution image of LMC-5 in I band taken by HST/WFPC2
six years after the lensing (Alcock \etal 2001).  The lens is the object
in the far left, and is $\sim3$ PC pixels  (0.134") away from the source
which is dimmer than the lens in $I$-band.  The bright object to the right
is a field star, and is 1.90 arcsec from the source.  (c), (d), (e) and
(f) show  LMC-5 in IRAC's 3.6, 4.5, 5.8 and 8$\micron$ bands.  
Note that the field star is
brighter than LMC-5 in $R$ and $I$ bands but comparable or dimmer in all
IRAC bands, indicating that IR flux is dominated by the cooler lens
which is moving away from source at roughly 24 mas per year. 
(a), (c), (d), (e) and (f) have the same plate scale. In
all images, North is up and East is to the left.}}
\end{figure}

\acknowledgements
This work is based (in part) on observations made with the Spitzer Space
Telescope, which is operated by the Jet Propulsion Laboratory, California
Institute of Technology under NASA contract 1407. Support for this work
was provided by NASA through an award issued by JPL/Caltech. 
Support for the IRAC instrument was provided by NASA through Contract
Number 960541 issued by JPL.  H. Nguyen 
thanks A. Goldin for his assistance. The authors would like to 
thank M.L.N. Ashby for helpful comments on the draft of the paper.

\end{document}